\documentclass{imammb}

\jno{dqnxxx}

\usepackage{natbib}
\setcitestyle{authoryear,open={(},close={)}}

\usepackage{graphicx}
\usepackage{amsmath,amsthm,bm,mathrsfs}
\usepackage{amssymb}
\usepackage{subcaption, floatrow}
\usepackage[backref=true,pagebackref=true,hyperindex=true,breaklinks=true,colorlinks=true,urlcolor=green,bookmarks=true,bookmarksopen=false,citecolor=blue,linkcolor=blue]{hyperref}
\usepackage{doi}
\usepackage[percent]{overpic}
\usepackage{multirow}
\usepackage{xcolor}
\usepackage{cancel}

\newcommand{\hide}[1]{}

\newcommand{\rev}[1]{#1}

\newcommand{\Vm}{V_\text{m}}
\newcommand{\Vf}{V_\text{f}}
\newcommand{\Va}{V_\text{m}^\alpha}
\newcommand{\Vfa}{V_\text{f}^\alpha}

\newcommand{\Gg}{G_\text{gap}}
\newcommand{\Gm}{G_\text{m}}
\newcommand{\Gf}{G_\text{f}}

\newcommand{\+}[3]{\def#1{{#2}}}
\+{\Eh}{\E_h}{h-gate switch voltage}
\+{\Em}{\E_m}{m-gate switch voltage}
\+{\E}{E}{transmembrane voltage}
\+{\F}{F}{an arbirary function}
\+{\GG}{G}{Inverse of $\FF$}
\+{\N}{N}{number of abstract dyn eqns}
\+{\eps}{\epsilon}{the small parameter of the main embedding}
\+{\h}{h}{fast inward current inactivation gate}
\+{\jbar}{\Qstat{j}}{fast inward current slow inactivation gate quasistat}
\+{\l}{l}{summation index in various places}
\+{\m}{m}{fast inward current activation gate}
\+{\taul}{\tau_{\l}}{relaxation timescale of $\l$-th var}
\+{\t}{t}{original (slow) time}
\+{\w}{w}{abstract dynamic variable; aux var \eq{change-to-bessel}}

\numberwithin{equation}{section}

\allowdisplaybreaks
\begin{document}

\title{Addendum: Action potential propagation and block in a model
of atrial tissue with myocyte-fibroblast coupling} 
\author{{\sc Peter Mortensen}\\[2pt]
School of Mathematics \& Statistics, University of Glasgow, Glasgow G12 8QQ\\
Institute of Cardiovascular \& Medical Sciences, University of Glasgow, Glasgow G12 8TA\\[6pt]
{\sc Hao Gao}\\[2pt]
School of Mathematics \& Statistics, University of Glasgow, Glasgow
G12 8QQ\\[6pt]
{\sc Godfrey Smith}\\[2pt]
Institute of Cardiovascular \& Medical Sciences, University of Glasgow, Glasgow G12 8TA\\[6pt]
{\sc and}\\[6pt]
{\sc Radostin D.~Simitev$^\ast$}\\[2pt]
School of Mathematics \& Statistics, University of Glasgow, Glasgow
G12 8QQ\\
$^\ast${\rm Corresponding author. Email:
  \href{mailto:Radostin.Simitev@glasgow.ac.uk}{Radostin.Simitev@glasgow.ac.uk}}\\[8pt]
{\rm [Received on 23 February 2021; revised on 26 March 2021; accepted on 14 April 2021]}\\[7pt]}
\pagestyle{headings}
\maketitle
\markboth{\rm P.~MORTENSEN \textit{ET AL.}}{\rm ADDENDUM: PROPAGATION \& BLOCK IN A
MYOCYTE-FIBROBLAST MODEL}


\begin{abstract}
{
The analytical theory of our earlier study (Mortensen et al. (2021),
\textit{Mathematical Medicine and Biology}, \rev{38(1), pp.~106-131}) is extended to address the
outstanding cases of fibroblast barrier distribution and myocyte
strait distribution.
In particular, closed-form approximations to the resting
membrane potential and to the critical parameter values for
propagation  are derived for these two non-uniform fibroblast
distributions and are in good agreement with numerical estimates.
}{electrophysiology, myocyte---fibroblast coupling, 
  refractoriness, asymptotic approximation}
\end{abstract}

\section{Introduction}
\begin{figure*}[!t]
\begin{center}
\includegraphics[width=0.7\linewidth]{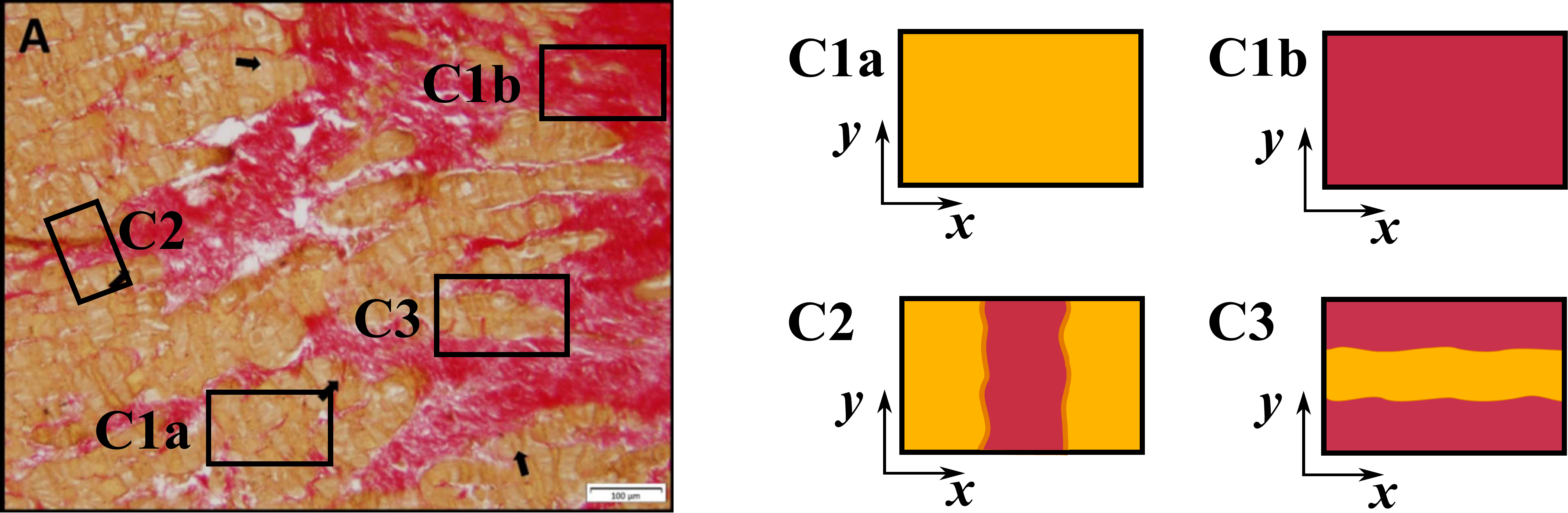}
\end{center}
\caption{\looseness=-1 \rev{A  schematic illustration 
of fibroblast distributions C1 (uniform), C2 (fibroblast barrier) and
C3 (myocyte strait) as they may appear in a border zone between intact
myocardium  and fibrosis.  Red shades indicate high
fibroblast density. To the left is an image of a border zone adapted by
permission from BMJ Publishing Group
\citep{Wald2018}. Propagation fronts are considered plane waves locally
so coordinate systems are attached to the idealised
regions only and propagation is along the $x$ axis.}  
}
\label{fig:FibrosisLabel}
\end{figure*}

The spatial distribution of fibroblasts within myocardial tissue, and
the electrical coupling between fibroblasts and cardiomyocytes are
significant but poorly-understood factors
in triggering and sustaining cardiac arrhythmias.  
In our recent article \citep{Mortensen2021} entitled ``Action
potential propagation and block in a model of atrial tissue with
myocyte-fibroblast coupling'', hereinafter referred to as MGSS, we
proposed that electrical propagation in fibrous tissue may be
understood conceptually on the basis of three elementary fibroblast
distributions: (C1) uniform, (C2) fibroblast barrier, and (C3) myocyte
strait, \rev{illustrated in Fig.~\ref{fig:FibrosisLabel}}. Using
direct numerical simulations, we then estimated 
primary action potential biomarkers including conduction velocity,
peak potential and triangulation index and found that propagation
block occurs at certain critical values of the parameters defining
each of the elementary distributions.  Based on a fast-slow scale
analysis \citep{2006Simitev,Biktashev2008,Simitev2011}, we demonstrated
that the boundary of absolute refractoriness in myocyte-fibroblast
tissue is determined primarily by the value of the myocyte potential
ahead of a propagating pulse and used this to obtain a simple
analytic expression that captures with remarkable accuracy the block
of propagation in the case of a uniform fibroblast distribution (C1).

The purpose of this addendum is to extend the analytical theory of
MGSS to the remaining two elementary fibroblast distribution cases --
(C2) fibroblast barriers and (C3) myocyte straits.
In the following, we derive and report closed-form
approximations to the resting membrane potentials which allows us to
estimate, also in closed-form, the critical parameter
values separating electrical propagation from failure in these two
non-uniform distributions. Our motivation for this analysis, the
background literature on the topic and the notation used here are
identical to those introduced in MGSS unless explicitly stated
otherwise. Hence, we refer the reader to our earlier paper MGSS for
these and further details.

\section{Analytic approximations of propagation through fibroblast barriers
and myocyte straits}

Fibrous atrial tissue is modelled in MGSS as a 2D
continuum of atrial myocytes where a fixed number of identical
fibroblasts, $n(x, y)$, are connected in parallel at every Cartesian
point $(x,y)$ via an \rev{inter-cell conductance} $\Gg$. In particular, the
``fibroblast barrier'' distribution (C2) is defined by 
\begin{subequations}
\label{n}
\begin{gather} 
n(x,y)=N\, H\big(\Delta x/2 -x\big) H\big(x+\Delta x/2),
\label{nC2}
\end{gather}
and the ``myocyte strait'' distribution (C3) is defined by
\begin{gather}
n(x,y)=N\, \Big(H\big(-y-\Delta y/2)+ H\big(y-\Delta y/2)\Big),
\label{nC3}
\end{gather}
\end{subequations}
\rev{and these are illustrated in
  Fig.~\ref{fig:FibrosisLabel}}. \rev{Here $\Delta x$ and $\Delta y$
  are the widths of barriers and straits, respectively,  $N$ is a positive
  integer, and $H(\cdot)$ is the Heaviside step function.}
The propagation of electrical excitation in atrial tissues with such
distributions is then described by a boundary value problem for the
monodomain equations (2.1) of MGSS.
Further, in MGSS we demonstrated that the boundary of absolute
refractoriness is determined to a good approximation by the value of
\rev{the myocyte potential ahead of a propagating pulse (prefront
  voltage),} $\Va$. \rev{Note that, superscripts in $\Va$, and further
  below in $\Vm^0$, do not denote exponentiation.}

\begin{table}[t!]
{  \small
  \begin{tabular}{ llcccc  }
    \hline
\rule{0mm}{4mm} Case & Interval   &	$K_1$ & $K_2$ & $K_3$ & $\lambda$ \\[1mm]
   \hline 
\multirow{2}{17mm}{Fibroblast barrier \eqref{nC2}}
&\rule{0mm}{7mm}
$\displaystyle s\in\left[0,\frac{\Delta s}{2}\right]$
&$\displaystyle \frac{\sqrt{\kappa_1} \kappa_6}{2 (\kappa_3 \kappa_7 + \kappa_8 \sqrt{\kappa_1})}$
& $\displaystyle\frac{\sqrt{\kappa_1} \kappa_6}{2 (\kappa_3 \kappa_7 + \kappa_8 \sqrt{\kappa_1})}$
& $\displaystyle \frac{\kappa_4}{\kappa_3^2}$
& $\displaystyle \kappa_3 \kappa_5$
\\
&\rule{0mm}{7mm}
$\displaystyle s\in\left[\frac{\Delta s}{2},\infty\right)$
  & $\displaystyle -\frac{\kappa_3 \kappa_6 \kappa_7}{\kappa_9(\kappa_3 \kappa_7  + \kappa_8 \sqrt{\kappa_1})}$
  & $\displaystyle 0$
  & $\displaystyle \Vm^0$
  & $\displaystyle \sqrt{\kappa_1}\kappa_5$
  \rule[-4ex]{0pt}{0pt}  \\ \hline
\multirow{2}{17mm}{Myocyte strait \eqref{nC3}}
&\rule{0mm}{7mm}   $\displaystyle s\in\left[0,\frac{\Delta s}{2}\right]$
&$\displaystyle \frac{-\kappa_3 \kappa_6}{2 (\kappa_3 \kappa_{11} + \kappa_{10} \sqrt{\kappa_1})}$
& $\displaystyle \frac{-\kappa_3 \kappa_6}{2 (\kappa_3 \kappa_{11} + \kappa_{10} \sqrt{\kappa_1})}$
& $\displaystyle \Vm^0$
& $\displaystyle \sqrt{\kappa_1}\kappa_5$
\\
&\rule{0mm}{7mm}~$\displaystyle s\in\left[\frac{\Delta s}{2},\infty\right)$
  & $\displaystyle -\frac{\sqrt{\kappa_1} \kappa_6 \kappa_{10}}{\kappa_{12}(\kappa_3 \kappa_{11}  + \kappa_{10} \sqrt{\kappa_1})}$
  & $\displaystyle 0$
  & $\displaystyle \frac{\kappa_4}{\kappa_3^2}$
  & $\displaystyle \kappa_3 \kappa_5$
\rule[-4ex]{0pt}{0pt}  \\ \hline
\end{tabular}}
  \caption{Exponentials and constants of integration in the 
    solution \eqref{equilsolution} of equations \eqref{EQ:problem}.}	
\label{t:const}
\end{table}

\looseness=-1
We now consider action potentials travelling \rev{along the $x$ axis} in fully rested
tissue with fibroblast distributions \eqref{n}. Then, the \rev{prefront
voltage} $\Va$ is identical to the steady state myocyte potential which
we proceed to determine.
Distributions \eqref{n} are functions of a
single variable denoted by $s$ for brevity, with $s$ being $x$ in the case of
a fibroblast barrier and $y$ in the case of a myocyte strait, respectively.
Since these functions are also even, the steady state version of
equations (2.1) of MGSS can be reduced in both fibroblast
distribution cases to a one-dimensional, time-independent
system posed on the real half-line   
\begin{subequations}
\label{EQ:problem}
\begin{gather}
\label{EQ:problem:Vm}  
\frac{d^2}{ds^2} \Va =  \frac{\chi}{\sigma_{ss}}\left(\Gm(\Va-\Vm^0) + n(s)
\Gg(\Va - \Vfa)\right), ~~~~ s\in\mathbb{R}_+,\\
\label{EQ:problem:Vf}  
0 = \Gf (\Vf^\alpha-\Vf^0) + \Gg (\Vfa - \Va),\\
\label{EQ:problem:BC}  
\frac{d}{ds}\Va(0)=0, \quad  \left[\frac{d}{ds}\Va\right]_{s\to\infty}=0.
\end{gather}
\end{subequations}
\rev{Here, $\chi$ is the cell
surface-to-volume ratio, $\sigma_{ss}$ with $s=x,y$ are the relevant
diagonal component of the conductivity tensor and their values are listed in
Table 1 of MGSS.}
To make the problem analytically tractable, we have also
linearized the equations near the uncoupled resting 
potentials $\Vm^0= -81$ mV and $\Vf^0= -46$ mV of the original
myocyte and fibroblast models of \cite{CRN98} and \cite{Morgan2016} as
detailed in section 4.5 of MGSS, \rev{with $\Gm$ and $\Gf$ being 
the coefficients to the leading-order Taylor series terms.}
Equation \eqref{EQ:problem:Vf} is easily solved for \rev{the fibroblast
resting potential $\Vfa$}, and we are left with a boundary-value
problem for a single second-order linear inhomogeneous ordinary
differential equation for \rev{the myocyte resting potential $\Va$}. Since the distributions $n(s)$
are piecewise constant, this resulting equation has constant coefficients in
each of the intervals $s\in[0,\Delta s/2]$ and $s\in[\Delta
s/2,\infty)$. Imposing continuity and smoothness matching conditions
at $s=\Delta s/2$ and boundary conditions \eqref{EQ:problem:BC}, we find
the solution
\begin{subequations}
\label{equilsolution}
\begin{gather}
  \Va(s) = K_1\exp(-\lambda s)+K_2\exp(\lambda s)+K_3,\\
  \Vfa(s) ={\big(\kappa_2 \Va(s) + \Vf^{0}\big)}/{\big(\kappa_2 + 1\big)}.
\end{gather}
\end{subequations}
Here the constants $K_1$, $K_2$, $K_3$ and $\lambda$ take different
values in the intervals $s\in[0,\Delta s/2]$ and $s\in[\Delta
s/2,\infty)$ and for the different choices of fibroblast distributions
\eqref{n} as listed in Table \ref{t:const} with further notation defined as
\begin{gather}
    \kappa_1=\frac{\Gm}{\Gg}, ~~
    \kappa_2=\frac{\Gf}{\Gg}, ~~
    \kappa_3^2=\frac{N \kappa_2 +   \kappa_1 \kappa_2 + \kappa_1}{\kappa_2 + 1}, ~~
    \kappa_4=\frac{ N \kappa_2 \Vf^{0} + \kappa_1\Vm^{0}(\kappa_2+1)}{\kappa_2 + 1},~~
    \kappa_5= \sqrt{\frac{\chi\Gg}{\sigma_{ss}}},\nonumber\\
      \label{terms}
    \kappa_6 =\Vm^0 - \frac{\kappa_4}{\kappa_3^2}, ~~
    \kappa_7 =\sinh\left(\kappa_3 \kappa_5\frac{\Delta s}{2}\right), ~~
    \kappa_8 =\cosh\left(\kappa_3 \kappa_5\frac{\Delta s}{2}\right), ~~
    \kappa_9=\exp\left(-\sqrt{\kappa_1}\kappa_5\frac{\Delta s}{2}\right),\\
    \kappa_{10} =\sinh\left(\sqrt{\kappa_1} \kappa_5\frac{\Delta s}{2}\right), ~~
    \kappa_{11} =\cosh\left(\sqrt{\kappa_1} \kappa_5\frac{\Delta s}{2}\right), ~~
    \kappa_{12}=\exp\left(-\kappa_3\kappa_5\frac{\Delta s}{2}\right).\nonumber
  \end{gather}

\begin{figure*}[!t]
\raisebox{-0.5mm}{\begin{overpic}[width=0.51\textwidth,height=0.346\textwidth]{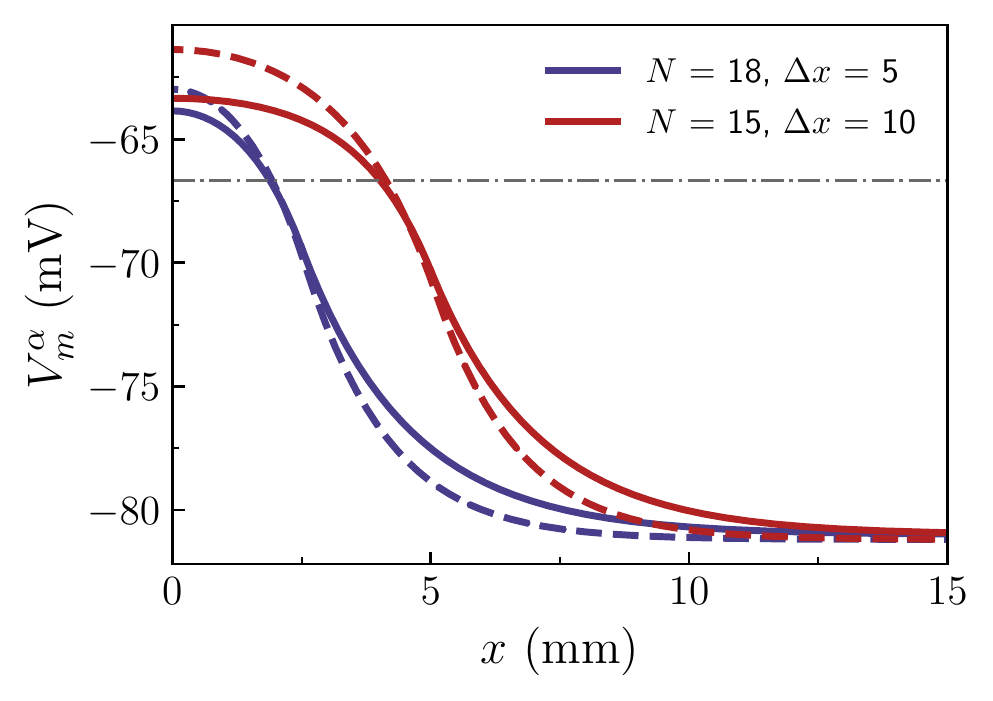}
  \put (2,62) {{(a)}}
\end{overpic}}
\begin{overpic}[width=0.48\textwidth,height=0.342\textwidth]{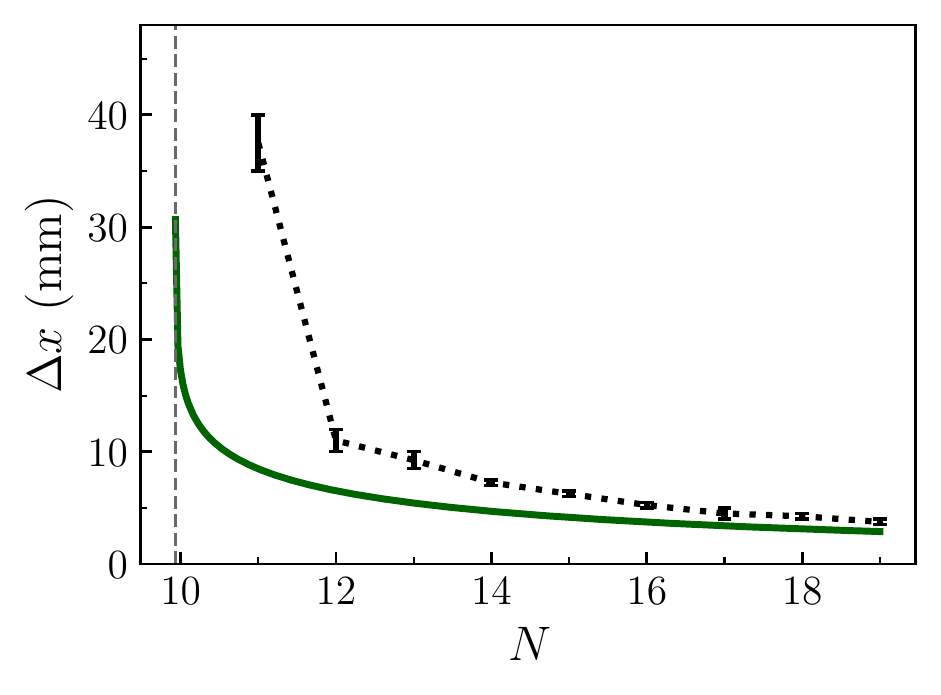}
\put (2,65) {{(b)}}                                          
\end{overpic}
\caption{
Closed-form approximations in comparison to direct numerical
simulations of MGSS for the case of fibroblast
barrier distribution \eqref{nC2}.
(a) Resting myocyte potential $\Va$. Expression \eqref{equilsolution}
is shown by solid curves and numerical results are shown by broken
curves both at values of $N$ and $\Delta x$ given in the legend. The
thin dash-dotted line is the line $\Va=\Eh$. \rev{Propagation if
successful is along the $x$ axis.} (b) Critical curve $\Delta
x(N)$. Expression \eqref{EQ:critC2} is shown by a solid green curve and 
numerical results are shown by a dotted black curve with error bars.
\rev{The numerical curve with error bars is the same one shown in 
Figure 6(a) of MGSS.}
The vertical asymptote $N_\text{asy}$ is shown by thin dashed line.
\label{fig:Vm0Case2}
}   
\end{figure*}

The critical curves $\Delta s(N)$ partitioning the \rev{generic} parameter planes
$\pi(N,\Delta s)$ of the problem into regions of propagation and
no-propagation are determined from the conditions
\begin{gather}
\label{eq:block}
\max_{s\in[0,\infty)}  \Va(s)=\Eh, ~~~\text{and}~~~ \min_{s\in[0,\infty)}  \Va(s)=\Eh,
\end{gather}
for fibroblast barrier and myocyte strait cases, respectively, as
detailed in the formulation of equation (4.12) of MGSS.
By the first of boundary conditions \eqref{EQ:problem:BC}, the profile
$\Va(s)$ has an absolute extremum at $s=0$ equal 
to ${\Va}(s=0) =2K_1+K_3$.
Hence, in the case of the fibroblast barrier distribution \eqref{nC2},
condition \eqref{eq:block} becomes
\begin{align}
\label{critcurveeqn}
\frac{\sqrt{\kappa_1} \kappa_6}{(\kappa_3 \kappa_7 + \kappa_8 \sqrt{\rev{\kappa_{1}}})}
+\frac{\kappa_4}{\kappa_3^2} &=\Eh.
\end{align}
This equation can be solved exactly for $\Delta s$. Indeed, the only
terms that depend on $\Delta s$ are $\kappa_7$ and $\kappa_8$, and
noting from \eqref{terms} that $\kappa_7=(\kappa_8^2-1)^{1/2}$, we solve
the equation for $\kappa_8$. Using the definition of $\kappa_8$, we
find a closed-form approximation for the critical curve in the case of
the fibroblast barrier distribution \eqref{nC2}
\begin{gather}
\label{EQ:critC2}
\Delta x = \frac{2}{\kappa_3\kappa_5} \cosh^{-1}
\left(\frac{\kappa_3 \sqrt{ (\Eh-\kappa_4/\kappa_3^2)^2(\kappa_3^{2} - \kappa_1) + \kappa_6^2 \kappa_1} + \kappa_6 \kappa_1}{|\Eh-\kappa_4/\kappa_3^2|(\kappa_3^{2} -  \kappa_1)}\right).
\end{gather}
Since $K_1$ does not vanish, we note that equation
\eqref{critcurveeqn} cannot have any solutions precisely when 
$\Eh=K_3$.
The latter defines a vertical asymptote for the critical curve and can
be solved explicitly to find 
\begin{gather}
N_\text{asy}=\frac{\kappa_1 (\kappa_2+1)(\Vm^0- \Eh)}{\kappa_2 \left(\Eh - \Vf^{0}\right)}.
\end{gather}
This is in agreement with equation (4.12) of MGSS for the uniform
distribution (C1) as the latter can be thought of as an infinitely wide fibroblast barrier.

\begin{figure*}[!t]
\raisebox{-0.5mm}{\begin{overpic}[width=0.50\textwidth,height=0.346\textwidth]{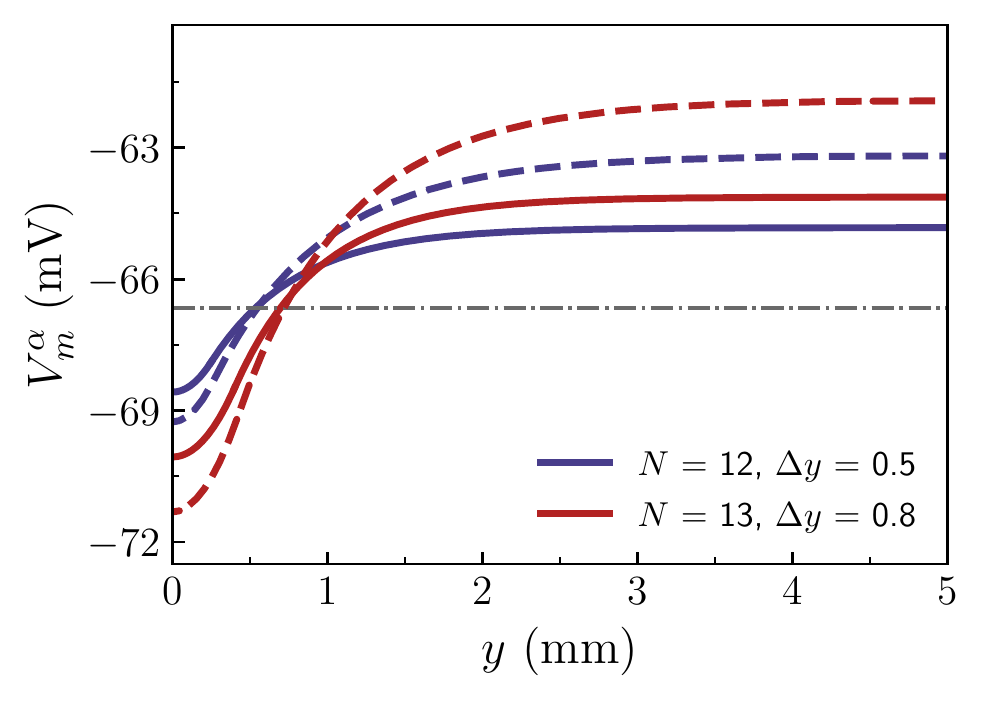}
  \put (2,64) {{(a)}}
\end{overpic}}
\begin{overpic}[width=0.485\textwidth,height=0.342\textwidth]{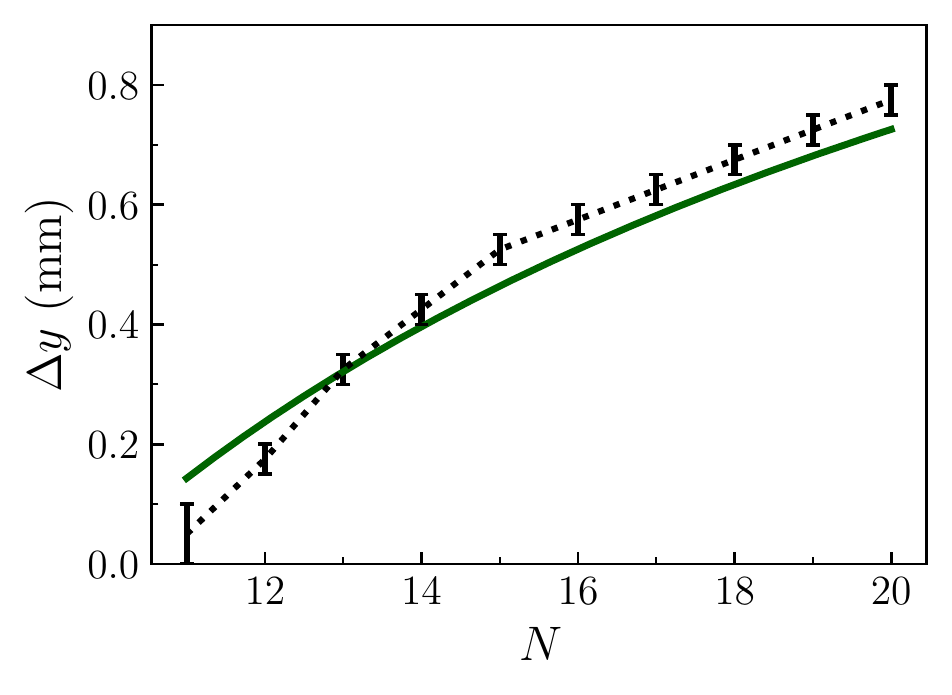}
\put (2,65) {{(b)}}                                          
\end{overpic}
\caption{
Closed-form approximations in comparison to direct numerical
simulations of MGSS for the case of myocyte strait
distribution \eqref{nC3}.
(a) Resting myocyte potential $\Va$. Expression \eqref{equilsolution}
is shown by solid curves and numerical results are shown by broken
curves both at values of $N$ and $\Delta y$ given in the legend. The
thin dash-dotted line is the line $\Va=\Eh$.
\rev{Propagation if successful is perpendicular to the $y$ axis.}
(b) Critical curve $\Delta
y(N)$. Expression \eqref{EQ:critC3} is shown by a solid green curve and 
numerical results are shown by a dotted black curve with error bars.
\rev{The numerical curve with error bars is the same one shown in Figure 8(a) of MGSS.}
\label{fig:Vm0Case3}
}   
\end{figure*}

For the myocyte strait distribution \eqref{nC3} the
approximation for the critical curve takes the form
\begin{gather}
\label{EQ:critC3}
\Delta s = \frac{2}{\sqrt{\kappa_1}\kappa_5} \cosh^{-1}
\left(
\frac{\kappa_{3}^{2} \kappa_{6} + \sqrt{\kappa_{1} \big(\kappa_{3}^{2} \kappa_{6}^{2} + (\Eh - \Vm^{0})^{2} \left(\kappa_{1} - \kappa_{3}^{2}\big)\right)}}{\left(\Eh - \Vm^{0}\right) \left(\kappa_{1} - \kappa_{3}^{2}\right)}
\right).
\end{gather}
This is obtained from the condition $\min_{s\in[0,\infty)}
\Va(s)=2K_1+K_3=\Eh$ in a similar way to the fibroblast barrier case
but rewriting first in terms of $\kappa_{11}$ rather than
$\kappa_8$. This curve does not have a vertical asymptote as
$\Vm^0\ne\Eh$. 

Figures \ref{fig:Vm0Case2} and \ref{fig:Vm0Case3} show a comparison of
the derived closed-form approximations with values from the direct
numerical simulations reported in MGSS for both fibroblast
distributions.
The discrepancy between the expression for the resting myocyte
potential \eqref{equilsolution} and the numerical values is due to
retaining only the linear terms in the Taylor expansions
leading to the right-hand sides of the first two of equations 
\eqref{EQ:problem}.
\rev{The expressions for the critical curves \eqref{EQ:critC2} and
\eqref{EQ:critC3} are compared to the corresponding numerical curves
shown in Figures 6(a) and 8(a) of MGSS, respectively, and} their
accuracy is additionally affected by the asymptotic reduction 
procedure used in MGSS to separate the description of fronts from the
description of steady state equilibrium.  However, the linearisation
errors and the asymptotic reduction errors seem to compensate each other
resulting in a remarkably close agreement 
between the analytic and the numerical results for the critical curves
of propagation. The direct numerical simulations of MGSS are, of
course, also subject to numerical errors which are harder to estimate.

\section{Conclusion}
Two archetypal non-uniform spatial distributions of
myocyte-fibroblast coupling were considered here. Approximations to
the resting potentials of the coupled cells and to the distribution
parameters at which action potential propagation is blocked were
derived in explicit analytic form and are in good 
correspondence with values from direct numerical simulations.
The results of the addendum are significant as they provide
theoretical underpinning of realistic 2D and 3D 
computational studies where high fibroblast density as opposed to
collagen accumulation leads to resting depolarization 
and spatial distribution of refractoriness \citep{McDowell2011} and to the
generation of complex fractionated atrial electrograms \citep{Ashihara2012}. 
Further, closed-form approximations of propagation in inhomogeneous
medium such as the ones derived here can be used to estimate poorly
constrained values of histological and electrophysiological parameters of myocardial tissue. 

\paragraph{Acknowledgements}
This work was supported by the UK Engineering and Physical Sciences
Research Council [grant numbers EP/N014642/1, EP/S030875/1 and EP/T017899/1]. 

\setlength{\bibsep}{1.0mm}

\end{document}